\documentstyle[12pt,aps,prd,epsf,amsfonts]{revtex}

\newcommand{\be}{\begin{equation}}
\newcommand{\ee}{\end{equation}}
\newcommand{\bea}{\begin{eqnarray}}
\newcommand{\eea}{\end{eqnarray}}

\def\rp{$R_p \hspace{-1em}/\ \  $}

\begin{document}
\draft
\title{Can LSND be included in a 3-Neutrino framework?}
\author{O. Haug $^{1}$, Amand Faessler$^1$ and J. D. Vergados$^{1,2}$}
\address{$^1$Institut f\"ur Theoretische Physik, Universit\"at T\"ubingen,
Auf der Morgenstelle 14, D-72076 T\"ubingen, Germany}
\address{
$^2$Theoretical Physics Division, Ioannina University, Ioannina, Greece}
\date{\today}
\maketitle
\begin{abstract}
We study the special features emerging from a three lepton generation
analysis of the available neutrino oscillation data (solar, atmospheric
and LSND). First we find that it is possible to explain all three sets of
data in terms of the standard left handed neutrinos without the need of
sterile neutrinos. Second we find a significant difference in the
mass matrix extracted from the data, depending on the analysis (without
or with LSND), if the mass of the lightest neutrino, which cannot be
determined from the neutrino oscillation data alone, is relatively
small, i.e $\leq$0.1 eV.
To compare with other processes we used the 
R-parity violating Minimal Supersymmetric Standard Model
(\rp-MSSM) for the 
theoretical description 
of the neutrino masses.
Using the oscillation data we were able to constrain the parameters of
the model. In particular we were able to obtain values for the 
coupling constants of the \rp-MSSM.
\end{abstract}
\pacs{PACS number(s):\{\}}
\renewcommand{\theequation}{\arabic{equation}}\setcounter{equation}{0}
Searches for neutrino masses have intensively been performed during the 
last decades. But till now no evidence for massive neutrinos was found in
experiments aiming to measure the mass directly, 
but only upper limits were set as follows
\cite{Caso:1998tx}
\bea
m_{\nu_{e}}<3.5 eV, & m_{\nu_{\mu}} < 160 KeV, & m_{\nu_{\tau}}<23MeV.
\eea 
On the other hand
some hints for massive neutrinos have been seen more than 30 years ago
in experiments measuring the electron neutrino flux coming from the sun 
\cite{Davis:1968cp}.
In these experiments
the flux of $\nu_e$ coming from the sun to the earth was found to be much 
smaller than the expectations. This observed difference of the expected 
vs the measured flux could be explained in 
terms of neutrino oscillations. The $\nu_e$ mixing, 
 due to oscillations to other neutrino flavors, 
leads to a smaller number of $\nu_e$'s at the detector.
These oscillations between 
different states can only occur if the weak eigenstates are different 
from the 
mass eigenstates and if the mass eigenstates are not all degenerate. 
This means that some neutrinos must be massive. The probability for a 
$\nu $, which 
was produced in the flavor state $|\nu_{\alpha} \rangle $ with energy 
$E$, to be detected in the flavor 
state $|\nu_{\beta}\rangle $ is given in a three family mixing scheme by
\bea
P(\alpha \rightarrow \beta) 
& = &
\delta_{\alpha , \beta} - 4 \sum_{i < j = 1}^{3}
U_{\alpha i} U_{\beta i} U_{\alpha j} U_{\beta j}
\sin^2
\left[
\frac{\Delta m^2_{ij} L}{4 E}
\right],
\label{prob}
\eea
where L is the source detector distance.
Here $\Delta m^2_{ij} \equiv \left| m_i^2 - m_j^2 \right| $ 
is the difference of the 
squared masses of the mass eigenstates $i$ and $j$, $U_{\alpha i}$ are the 
elements $\alpha i$ of the mixing matrix $U$,
assuming it is real (negligible CP violation),
which connects the flavor 
eigenstates $|\nu_{\alpha} \rangle $ and the mass eigenstates 
$|\nu_{i}\rangle $
by $|\nu_{\alpha} \rangle = \sum_i U_{\alpha i}|\nu_i \rangle $. 
If CP-violation is 
negligible the mixing matrix $U$ can be parameterized in analogy to the CKM 
matrix by three angles,
\begin{eqnarray}
U & = & 
\left(
\begin{array}{ccc}
c_{12} c_{13} & s_{12} c_{13} & s_{13} \\
-s_{12} c_{23}-c_{12}s_{23}s_{13} & c_{12}c_{23}-s_{12}s_{23}s_{13}&
s_{23}c_{13} \\
s_{12}s_{23}-c_{12}c_{23}s_{13} & - c_{12}s_{23}-s_{12}c_{23}s_{13}&
c_{23}c_{13}
\end{array}
\right),
\end{eqnarray}
where $s_{ij}$ and $c_{ij}$ stand for $\sin ( \theta_{ij} )$ and 
$\cos ( \theta_{ij} )$, respectively.

The search for neutrino oscillations 
has been the subject of many recent experiments 
\cite{zeitnitz:98,homestake:98,sage:98,Casper:1991ac,fukuda:98,Fukuda:1999pp,lsnd:97,lsnd:98}.
The experiments which found evidence for the existence of $\nu $ oscillations 
can be characterized by the sources of neutrinos which they use as 
solar \cite{homestake:98,sage:98}, 
atmospheric \cite{Casper:1991ac,fukuda:98,Fukuda:1999pp} 
and accelerator neutrino experiments \cite{lsnd:97,lsnd:98}. 
For the atmospheric and solar neutrino experiments several groups find 
evidence for neutrino oscillations. In the field of accelerator neutrino 
experiments only one group, the LSND collaboration, has claimed evidence
for neutrino oscillations, but this result has not yet been confirmed 
by any other experiment.
Furthermore it is commonly believed that all the three experiments
(solar, atmospheric and LSND) cannot be accommodated with just two
independent $\Delta m^2$. Thus the LSND results were put in doubt.
We will see that this outcome is due to the essentially
two flavor scenario, which has hitherto been employed in most analyses.
 We will see that in a rigorous three generation analysis all the
available data can be accommodated and can be used to constrain the
neutrino mass matrix.
Since the emerging solution is, unfortunately, not unique,
we propose a way 
to test in the future this solution and thus, indirectly, to test the 
LSND results.

As we have already mentioned, the
analysis of neutrino oscillation experiments is often done 
independently for each 
experiment within a two family scenario. The 
oscillation probability with the same notation as in eqn. (\ref{prob}) in
a two generation mixing scheme is
\bea
P(\alpha \rightarrow \beta) 
& = &
\sin^22 \theta \sin^2
\left(
1.27 \Delta m^2 \frac{L}{E}
\right),
\label{prob2}
\eea
where $\Delta m^2$ is in Units eV$^2$, $L$ in $m$ and $E$ in MeV.
The result of analyses of the solar neutrino experiments for the mass 
squared difference $\Delta m^2$ by including 
matter effects 
(Mikheyev-Smirnov-Wolfenstein\cite{Wolfenstein:1978ue,Mikheev:1985gs}) is
\bea
4\times 10^{-6} eV^2 &\leq \Delta m^2_{sun} \leq 1.2 \times 10^{-5} eV^2.
\label{solar2}
\eea
For the atmospheric $\nu$ experiments one finds 
disappearance oscillations of $\nu_{\mu}$ probably into $\nu_{\tau}$
and obtains for $\Delta m^2$
\bea
4 \times 10^{-4}eV^2 & < \Delta m^2 < 6 \times 10^{-3} eV^2.
\eea
Fitting now the results from the LSND collaboration for oscillations of 
$\bar{\nu}_{\mu}\rightarrow \bar{\nu}_e $ coming form $\mu^+$ decay at rest 
and $\nu_{\mu} \rightarrow \nu_e$ which originate form the decay of $\pi^+$ 
also in a two family analysis yields a 
mass splitting of
\bea
0.1 eV^2 < \Delta m^2 < 1.0 eV^2.
\label{lsnd2}
\eea
The three different splittings $\Delta m^2$ of 
eqns. (\ref{solar2}) to (\ref{lsnd2}) seem to be not 
compatible with 
a mixing of only three $\nu$ families and led to the conclusion that 
either the LSND result is wrong or that 
there should 
be a fourth  sterile neutrino which mixes with the others. But before 
one makes such severe claims
one should carefully scrutinize the methods which have been used to 
obtain these results. By looking at eqns. (\ref{prob}) and (\ref{prob2}) one 
realizes that the squared mass differences 
obtained by fits using a two family scenario and by using a three family 
scenario need not to be the same. Indeed the
squared mass difference extracted form eqn. (\ref{prob2}) 
are a convolution 
of the squared mass differences in the three family scheme.
Therefore the 
need for a sterile neutrino does not necessarily 
follow from the above argument. Thus one should analyze 
the experimental data not in the oversimplified 
two family but in a three family mixing scheme. This has been 
done by several authors \cite{Teshima:1998fi,Sakai:1999kc,scheck:98,thun:98,ohlsson:99,fogli:99,Fogli:1999fs}. 
The results of these analyses are displayed in 
table \ref{anareswithoutlsnd} and \ref{anareslsndincluded}.
In table \ref{anareswithoutlsnd} we show the results of 
such analyses which excluded 
the LSND results while in table \ref{anareslsndincluded} the results of 
analyses including the LSND results are given. 
At first we want to examine the influence of the inclusion of the 
LSND result on the oscillation parameters in the different analyses.
By comparing the two tables we 
see that the influence of the inclusion of the LSND experiment on the mixing 
angles $\theta_{13}$ and $\theta_{23}$ is negligible and is still relatively 
small for 
$\theta_{12}$. The mass splitting $\Delta m^2_{12}$ of the lower lying 
masses is found to be larger by including the LSND result 
in most analyses. Also for the mass 
splitting $\Delta m^2_{23}$ the inclusion of the LSND result leads to a 
larger splitting. 
In average, the inclusion 
of the LSND result in global fits to all data leads only to a larger
mass splitting $\Delta m^2_{ij}$, which is due to the small oscillation 
length ($L\simeq 20$m) in LSND.
To see now the consequences of this 
difference we calculated the entries of the mass matrix in the weak basis 
for the different analyses. Oscillations fix the difference of the masses 
squared but not the absolute scale of the masses.
For values of the smallest mass $m_1$ larger than 1 eV we find no 
significant difference between the two types of analyses.
For three cases of the smallest mass, 
$m_1=0.0,$ 0.01 and 0.1 eV, we show the average value 
for the neutrino mass matrix for the two types of analyses
in table \ref{xxx}. By comparing the results one finds that there exists a 
significant difference between the two types of solutions for these 
small masses. Because of the non 
observation of the neutrinoless double beta decay ($0\nu \beta \beta$) one 
expects such small masses for Majorana neutrinos 
\cite{faessler:98,simkovic:99}.
As it is well known the mass extracted from the $0\nu$ double beta
decay experiments is the absolute value of the quantity:
\bea
<m_{\nu}> &=&\sum_{j=1}^{3}U^2_{ej}~\lambda^{CP}_j~m_j
\label{bb1}
\eea
where $\lambda^{CP}_j$ are the CP eigenvalues of the neutrino mass
eigenstate $\nu_j$ with mass $m_j$, which, of course, is positive.
From the data of TABLE II we find:
\bea
<m_{\nu}> &=& 0.006,0.013,0.101,~~~~ \mbox{respectively}~~~~~(without~\mbox{LSND})
\label{bb2}
\eea
\bea
<m_{\nu}> &=& 0.026,0.032,0.115,~~~~\mbox{respectively}~~~~~(with~\mbox{LSND})
\label{bb3}
\eea
The small values can differentiate between the two types of analysis, 
but, unfortunately, they are far below the present experimental limits. 
They are, however, within the goals of the planned experiments
\cite{faessler:98,simkovic:99}.
The experimentally most interesting values obtained for large $m_1$ do
not seem to depend on the type of the analysis (with or without LSND).

To decide now which of the two kind of analyses is correct one 
has to check the consistency of the results with other 
physical observables. 
To compare with other processes we 
have to consider now a model which is able to accommodate
massive neutrinos. 
This means that we have to use a model which is an extension of the Standard 
Model $(SM)$. One natural way to extend the SM is supersymmetry $(SUSY)$
\cite{nilles:84}. 
A popular SUSY extension is the R-parity violating supersymmetric 
SM (\rp-$MSSM$). 
Within this model we have for each SM particle a supersymmetric 
partner, called sparticles. For details about the \rp-MSSM see e.g.
\cite{barbier:98}. 
Within this model all neutrinos acquire masses by mixing with the 
the SUSY partners of the photon, the $Z^0$ 
and the two neutral SUSY Higgs fields,
see fig. \ref{tree},
and by loop corrections, see fig. \ref{loops}.
For details how neutrinos 
acquire masses within this model see \cite{haug,bednyakov:98,rakshit:98}.
Using the values for the entries of the neutrino mass matrix 
given in table \ref{xxx} we 
extracted the values of the coupling constants for single term dominance. 
They are given in table \ref{limits}.
We see that for the couplings $\lambda^{(\prime)}_{i33}$
exists a significant difference only in the cases of 
$m_1=0.00$ and $m_1=0.01$ eV. These small masses are found in a
predictive model which use an underling symmetry to describe quark and 
lepton masses \cite{haug}. 

To conclude, we discussed the analyses of neutrino oscillation data and showed 
that it is necessary to use a three family mixing scheme to do such an 
analysis.
We presented results of different 
three family analyses including and excluding the LSND result and 
found that the inclusion mainly influences the mass splitting 
$\Delta m_{23}^2$. 
We found that to distinguish the two different approaches and 
test by this the LSND result 
will only be possible if the smallest mass is smaller than 0.01 eV.
\section*{Acknowledgments}
One of the authors (JDV) is happy to acknowledge partial support from
the TMR contract ERBFMRXCT96-0990.

\begin{table}[th]
\begin{tabular}{cccccc}
$\Delta m^2_{12}[eV^2]$&$\Delta m^2_{23}[eV^2]$&$\theta_{12}[Deg.]$
&$\theta_{23}[Deg.]$&$\theta_{13}[Deg.]$&ref.
\\
$3\times 10^{-6}-7 \times 10^{-5}$ & 0.01& 53-62 & 28-37 & $<$13 & 
\cite{Sakai:1999kc}\\
$4\times 10^{-6}-7 \times 10^{-5}$ & 1.0 & 51-72 & 27-32 & $<$4 & 
\cite{Teshima:1998fi}\\
$4\times 10^{-6}-7 \times 10^{-5}$ & 0.1 & 51-72 & 28-33 & $<$3 &
\cite{Teshima:1998fi}\\
$10^{-4}$ &$ 8\times 10^{-4}$ & 39.23 & 45 & 26.6  & \cite{fogli:99}\\
\end{tabular}
\caption{Results of neutrino oscillation analyses in a three family mixing 
scheme excluding the LSND results.}
\label{anareswithoutlsnd}
\end{table}
\begin{table}
\begin{tabular}{cccccc}
$\Delta m^2_{12}[eV^2]$&$\Delta m^2_{23}[eV^2]$&$\theta_{12}[Deg.]$
&$\theta_{23}[Deg.]$&$\theta_{13}[Deg.]$&ref.
\\
$10^{-4}-10^{-3}$ & 0.3& 35.5 & 27.3 & 13.1 & \cite{scheck:98}\\
$10^{-4}-10^{-3}$ & 0.3& 54.5 & 27.3 & 13.1 & \cite{scheck:98}\\
$2.87 \times 10^{-4} $ & 1.11 & 45 & 28.9 & 4.2 & \cite{ohlsson:99} \\
$10^{-4}-10^{-3} $ & 0.4 & 37.6 & 26.5 & 10.3 & \cite{thun:98} \\
$4\times 10^{-6}-7 \times 10^{-5}$ & 1 & 51-72 & 27-32 & 3-4 & 
\cite{Teshima:1998fi}\\
\end{tabular}
\caption{Results of neutrino oscillation analyses in three family mixing 
scheme including the LSND results.}
\label{anareslsndincluded}
\end{table}

\begin{table}
\begin{tabular}{c|ccc}
 & $m_1=0.00$ eV & $m_1=0.01$ eV & $m_1=0.10$ eV \\ \hline
without LSND & 
$
\left( 
\begin{array}{ccc}
.006 & .012 & .007 \\
.012 & .109 & .099 \\
.007 & .099 & .111 
\end{array}
\right)
$eV &
$
\left( 
\begin{array}{ccc}
.013 & .010 & .008 \\
.010 & .114 & .095 \\
.008 & .095 & .117 
\end{array}
\right)
$eV &
$
\left( 
\begin{array}{ccc}
.101 & .005 & .005 \\
.005 & .181 & .076 \\
.005 & .076 & .186 
\end{array}
\right)
$eV \\ \hline
LSND included & 
$
\left( 
\begin{array}{ccc}
.026 & .064 & .072 \\
.064 & .292 & .339 \\
.072 & .339 & .457 
\end{array}
\right)
$eV &
$
\left( 
\begin{array}{ccc}
.032 & .061 & .073 \\
.061 & .296 & .336 \\
.073 & .336 & .459 
\end{array}
\right)
$eV &
$
\left( 
\begin{array}{ccc}
.115 & .050 & .067 \\
.050 & .354 & .301 \\
.067 & .301 & .496 
\end{array}
\right)
$eV 
\end{tabular}
\caption{Results for the neutrino mass matrix in the weak basis. Given are 
the averaged values for matrix elements using the analyses discussed in tables 
\ref{anareswithoutlsnd} and \ref{anareslsndincluded} for $\lambda^{CP}_j$=1
which we found to give the largest differences. With LSND one 
obtains larger off-diagonal elements.
}
\label{xxx}
\end{table}
\begin{table}
\begin{tabular}{c|ccc|ccc}
&\multicolumn{3}{c|}{without LSND}&\multicolumn{3}{c}{LSND included}\\
$m_1$ $[eV]$ & 0.00 & 0.01 & 0.10 & 0.00 & 0.01 & 0.10 \\ \hline
$\frac{\lambda^{\prime}_{133}}{\sqrt{M_{SUSY}/100 GeV}}\times 10^{4}$&
0.4 & 0.6 & 1.6 & 0.8 & 0.9 & 1.7 \\
$\frac{\lambda^{\prime}_{233}}{\sqrt{M_{SUSY}/100 GeV}}\times 10^{4}$&
1.6 & 1.7 & 2.1 & 2.7 & 2.7 & 2.9 \\
$\frac{\lambda^{\prime}_{333}}{\sqrt{M_{SUSY}/100 GeV}}\times 10^{4}$&
1.6 & 1.7 & 2.1 & 3.3 & 3.3 & 3.5 \\
$\frac{\lambda_{133}}{\sqrt{M_{SUSY}/100 GeV}}\times 10^{4}$&
1.7 & 2.6 & 7.1 & 3.6 & 4.0 & 7.6 \\
$\frac{\lambda_{233}}{\sqrt{M_{SUSY}/100 GeV}}\times 10^{4}$&
7.4 & 7.5 & 9.5 & 12.1 & 12.2 & 13.3 \\
\end{tabular}
\caption{Values for trilinear couplings for single term dominance 
using neutrino oscillation analyses 
and different mass schemes for $m_1$.
For the calculation of these values of the trilinear coupling 
constants we assumed as usual that 
masses and soft parameters are of the same order and 
are parameterized by $M_{SUSY}$ which is expected to lie in the region from 
100 to 1000 GeV.}\label{limits}
\end{table}
\begin{figure}
\caption{Mixing of the neutrinos and neutralinos.}
\label{tree}
\end{figure}
\begin{figure}
\caption{Radiative contributions to the neutrino mass in the \rp-MSSM.}
\label{loops}
\end{figure}
\centerline{Fig. 1}
\vskip 1cm
\epsfysize=5cm 
\centerline{\epsffile{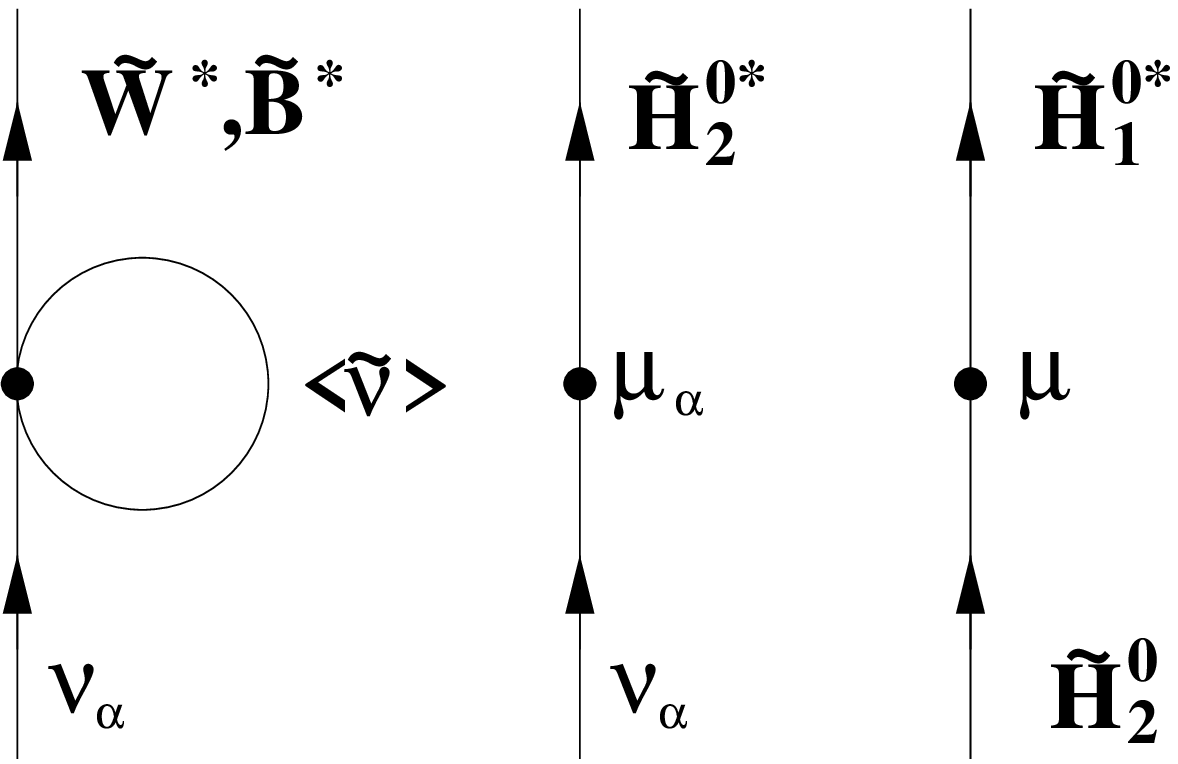}}
\centerline{Fig. 2}
\vskip 1cm
\epsfysize=5cm 
\centerline{\epsffile{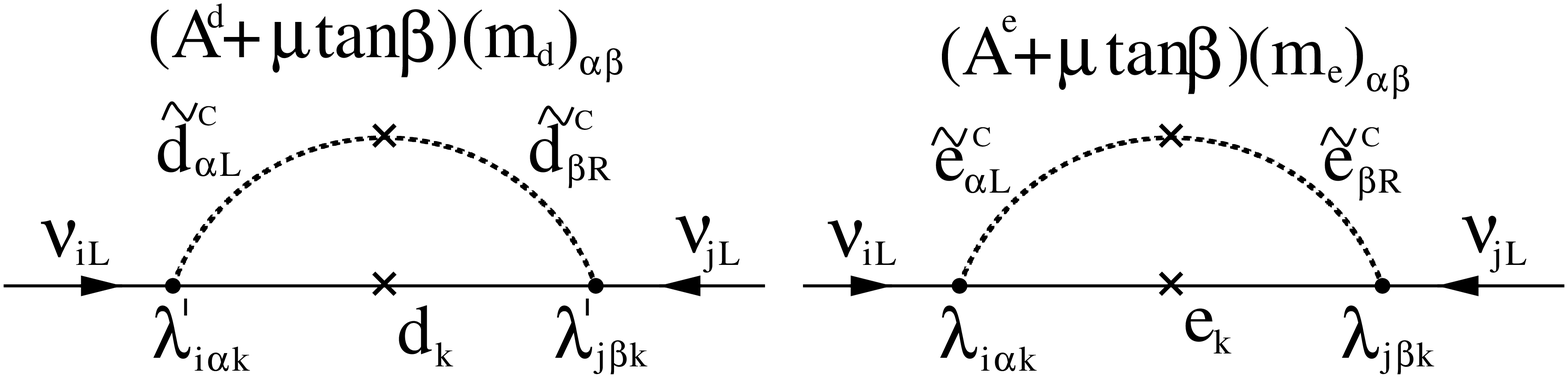}}
\end{document}